\documentclass{article}

\usepackage[english]{babel}
\usepackage{authblk}
\usepackage{float}

\usepackage{amsmath}
\usepackage{amssymb}
\usepackage{booktabs}

\usepackage[letterpaper,top=2cm,bottom=2cm,left=3cm,right=3cm,marginparwidth=1.75cm]{geometry}

\usepackage{amsmath}
\usepackage{graphicx}
\usepackage[colorlinks=true, allcolors=blue]{hyperref}

\title{\textbf{Sophisticated deep learning with on-chip optical diffractive tensor processing}}

\author{Yuyao Huang\textsuperscript{1}, Tingzhao Fu\textsuperscript{1}, Honghao Huang\textsuperscript{1}, Sigang Yang\textsuperscript{1}, and Hongwei Chen\textsuperscript{1,*}}

\affil[1]{Beijing National Research Center for Information Science and Technology, Department of Electronic Engineering, Tsinghua University, Beijing 100084, China}

\affil[*]{Corresponding author: chenhw@tsinghua.edu.cn}

\date{}

\begin{document}
\maketitle

\begin{abstract}
The ever-growing deep learning technologies are making revolutionary changes for modern life. However, conventional computing architectures are designed to process sequential and digital programs, being extremely burdened with performing massive parallel and adaptive deep learning applications. Photonic integrated circuits provide an efficient approach to mitigate bandwidth limitations and power-wall brought by its electronic counterparts, showing great potential in ultrafast and energy-free high-performance computation. Here, we propose an optical computing architecture enabled by on-chip diffraction to implement convolutional acceleration, termed optical convolution unit (OCU). We demonstrate that any real-valued convolution kernels can be exploited by OCU with a prominent computational throughput boosting via the concept of structral re-parameterization. With OCU as the fundamental unit, we build an optical convolutional neural network (oCNN) to implement two popular deep learning tasks: classification and regression. For classification, Fashion-MNIST and CIFAR-4 datasets are tested with accuracy of 91.63\% and 86.25\%, respectively. For regression, we build an optical denoising convolutional neural network (oDnCNN) to handle Gaussian noise in gray scale images with noise level \textbf{$\sigma = 10,15,20$}, resulting clean images with average PSNR of 31.70dB, 29.39dB and 27.72dB, respectively. The proposed OCU presents remarkable performance of low energy consumption and high information density due to its fully passive nature and compact footprint, providing a highly parallel while lightweight solution for future compute-in-memory architecture to handle high dimensional tensors in deep learning.
\end{abstract}

\section{Introduction}

Convolutional neural networks (CNNs) \cite{lecun1998gradient, lecun2015deep, gu2018recent} powers enormous applications in artificial intelligence (AI) world including computer vision \cite{krizhevsky2017imagenet, voulodimos2018deep, forsyth2002computer}, self-driving cars \cite{bojarski2016end,levinson2011towards,grigorescu2020survey} , natural language processing \cite{hirschberg2015advances,nadkarni2011natural,chowdhary2020natural}, medical science \cite{shen2017deep, gawehn2016deep,angermueller2016deep}, etc. Inspired by biological behaviors of visual cortex systems, CNNs have brought remarkable breakthroughs in manipulating high-dimensional tensor such as images, videos and speech, enabling efficient processing with more precise information extractions but much fewer network parameters, compared with the classical feed-forward one. However, advanced CNN algorithms have rigorous requirements on computing platforms who are responsible for massive data throughputs and computations, which triggers a flourishing development of high performance computing hardware such as central processing unit (CPU) \cite{hennessy2011computer}, graphics processing unit (GPU) \cite{kirk2007nvidia}, tensor processing unit (TPU) \cite{jouppi2017datacenter}, and field-programmable gate array (FPGA) \cite{zhang2015optimizing}. Nonetheless, today’s electronic computing architectures are facing physical bottlenecks in processing distribute and parallel tensor operations, mainly are bandwidth limitation, high-power consumption, and the fading of Moore’s Law, causing serious computation force mismatches between AI and the underlying hardware frameworks. 

Important progresses have been made to further improve the capabilities of future computing hardware. In recent years, optical neural network (ONN) \cite{caulfield2010future,miller2010role,touch2017optical,carleo2019machine,shastri2021photonics,bogaerts2020programmable,markovic2020physics,zhou2022photonic} receives growing attentions with its extraordinary performances in facilitating complex neuromorphic computations. The intrinsic parallelism nature of optics enables more than 10-THz interconnection bandwidth\cite{cheng2018recent}, and the analog fashion of photonics system \cite{yao2009microwave} decouplings the urgent needs for high-performance memory in conventional electronic architectures, and therefore prevents energy wasting and time latency from continuous AD/DA conversion and ALU-memory communication, boosting computational speed and reducing power consumption essentially. 

To date, numerous ONNs are proposed to apply various neuromorphic computations such as optical inference networks based on MZI mesh\cite{shenyc2017deep,wetzstein2020inference,zhang2021optical}, photonics spiking neural networks based on WDM protocol and ring modulator array \cite{tait2017neuromorphic,tait2014broadcast,huang2021silicon}, photonics tensor core based on phase change materials \cite{feldmann2021parallel,wu2021programmable}, optical accelerator based on time-wavelength interleaving \cite{huang2019programmable,xu2020photonic,xu202111}, etc. For higher computation capabilities of ONNs, diffractive optical neural networks \cite{lin2018all,zhou2021large,xu2022multichannel, doi:10.1126/sciadv.abn7630} are proposed to provide millions of trainable connections and neurons optically by means of light diffraction. In our previous work \cite{fu2021chip}, we demonstrated an integrated diffractive optical neural network in SOI platform to further improve network density with metasurface technologies. Here, we make one step forward to build an optical convolution unit (OCU) with the same method to achieve massive parallel tensor computations. We demonstrate that any real-valued
convolution kernels can be exploited by OCU with a prominent computation power. Furthermore, with OCU as the basic building block, we build an optical convolutional neural network (oCNN) to perform classification and regression tasks. For classification task, Fashion-MNIST and CIFAR-4 datasets are tested with accuracies of 91.63\% and 86.25\%, respectively. For regression task, we build an optical denoising convolutional neural network (oDnCNN) to handle Gaussian noise in gray scale images with noise level \textbf{$\sigma = 10,15,20$}, resulting clean images with average PSNR of 31.70dB, 29.39dB and 27.72dB. The proposed OCU and oCNN are fully passive in processing massive tensor data and compatible for ultrahigh bandwidth interfaces (for both electronic and optical), being capable of integrating with electronic processors to re-aggregate computational resources and power penalty.

\section{Principle}

Fig.1(a) presents the operation principle of 2D convolution. Here, a fixed kernel $\textbf{K}$ with size of $H\times H$ slides over the image $\textbf{I}$ with size of $N\times N$ by stride of $S$ and does weighted addition with the image patches that cover by the kernel, resulting an extracted feature map $\textbf{O}$ with size of $G\times G$, where $G = \lfloor (N - H)/S + 1 \rfloor$ (in this case we ignore the padding process of convolution). This process can be expressed in Eq.(1), where $\textbf{O}[i,j]$ represents a pixel of the feature map, $m$ and $n$ are related to the stride $S$. Based on this, one can simplify the operation as multiplications between a $H^{2} \times 1$ vector $\widehat{\textbf{K}}$ that reshaped by the kernel and a $G^{2} \times H^{2}$ matrix $\widehat{\textbf{I}}$ that composed by image patches, as shown in Eq.(2).

\begin{equation}
\textbf{O}[i,j]=\sum_{m=-\infty}^{\infty}\sum_{n=-\infty}^{\infty}\textbf{I}[i-m,j-n]\cdot\textbf{K}[m,n]
\end{equation}
\begin{equation}
\widehat{\textbf{O}}= \widehat{\textbf{K}} \cdot \widehat{\textbf{I}}=
\left[ \begin{array}{ccccc}
\emph{k}_{1} \\ \emph{k}_{2} \\ \cdots \\ \emph{k}_{\emph{H}^2}\\
\end{array} 
\right ]^{T}
\cdot
\left[ \begin{array}{ccccc}
\emph{i}_{11}  & \emph{i}_{12} & \cdots & \emph{i}_{1,\emph{G}^2}\\
\emph{i}_{21}  & \emph{i}_{22} & \cdots & \emph{i}_{2,\emph{G}^2}\\
\vdots & \vdots &\ddots & \vdots\\
\emph{i}_{\emph{H}^2,1} & \emph{i}_{\emph{G}^2,2} & \cdots & \emph{i}_{\emph{H}^2,\emph{G}^2}\\
\end{array} 
\right ]
\end{equation}

Here, $\widehat{\textbf{I}_{m}}=[\emph{i}_{m,1},\emph{i}_{m,2},...,\emph{i}_{m,H^2}]$ where $\emph{m}=(1,2,...,G^2)$ denotes a corresponding image patch covered by a sliding kernel and $\widehat{\textbf{O}}$ is the flattened feature vector. Consequently, the fundamental idea of optical 2D convolution is to manipulate multiple vector-vector multiplications optically and keep their products in series for reshaping to a new map. Here, we use on-chip optical diffraction to implement this process, as described in Fig.1(b). The input image with size of $\emph{N} \times \emph{N}$ is firstly reshaped into flattened patches according to the kernel size \emph{H} and sliding stride \emph{S}, which turns the image into a $\emph{G}^2 \times \emph{H}^2$ matrix $\widehat{\textbf{I}}$. Then $\widehat{\textbf{I}}$ is mapped into a plane of space channels and time, in which each row of $\widehat{\textbf{I}}$ is varied temporarily with period of $\emph{G}^2$ and each column of $\widehat{\textbf{I}}$ (namely pixels of a flattened image patch) are distributed in $\emph{H}^2$ channels. A coherent laser signal is split into $H^{2}$ paths and then modulated individually by the time-encoded and channel-distributed image patches, in either amplitude or phase. In this way, one time slot with duration of $\Delta \emph{t}$ contains one image patch with $H^{2}$ pixels in corresponding channels, and $G^{2}$ of these time slots can fully express image patch matrix $\widehat{\textbf{I}}$. Then the coded light is sent to the proposed OCU to perform matrix multiplications as Eq.(2) shows, and the corresponding positive and negative results are detected by a balanced photodetector (BPD) to do subtractions between the two. The balanced detection scheme assures OCU operates in real-valued field. The detected information is varied temporarily with symbol duration of $\Delta t$, and then reshaped into a new feature map by a digital signal processor (DSP). 

The details of OCU are given in Fig.1(c). Here, $H^{2}$ silicon strip waveguides are exploited for receiving signals simultaneously from modulation channels, who diffract and interfere with each other in a silicon slab waveguide with size of $L_1 \times L_2$ before it encounters well-designed 1D-metalines. The 1D-metaline is a sub-wavelength grating consists of silica slots with each slot has size of $w_1 \times w_2 \times h$, which is illustrated in the inset of Fig.1(c). Furthermore, we use 3 identical slots with slot gap of $\emph{g}$ to constitute a metaunit with period of $p$ to ensure a constant effective refractive index of the 1D-metaline when it meets light from different angles, as demonstrated in our previous work \cite{fu2021chip}. The incoming signal is phase-modulated from 0 to $2\pi$ by changing the length of each metaunit $w_2$ but with $w_1$ and $h$ fixed. Accordingly, the corresponding length $w^{(l)}_{2,v}$ of the \emph{v}-th metaunit in the \emph{l}-th metaline can be ascertained from the introduced phase delay $\Delta \phi^{(l)}_{v}$ by Eq.(3), where $\emph{n}_{1}$ and $\emph{n}_{2}$ are the effective refractive index of the slab and slots respectively. After layers of propagation, the interfered light is sent to two ports which outputs positive and negative part of computing results.

\begin{equation}
w^{(l)}_{2,v}=\frac{\lambda}{2\pi(n_1-n_2)}\cdot \Delta \phi^{(l)}_{v}
\end{equation}
\begin{equation}
\emph{f}^{\;(l)}_{u,v}=\frac{1}{\emph{j$\lambda$}}\cdot \Big(\frac{1+{\rm cos}\theta_{u,v}}{2r_{u,v}}\Big) \cdot {\rm exp}\Big(j\frac{2\pi r_{u,v}n_{1}}{\lambda}\Big)\cdot \eta {\rm exp}(j\Delta\psi)
\end{equation}

For a more precise analysis, the diffraction in the slab waveguide between two metalines with \emph{U} and \emph{V} metaunits respectively is characterized by a $\emph{U} \times \emph{V}$ matrix $\textbf{F}^{(l)}$ based on Huygens-Fresnel principle under restricted propagation conditions, whose element $\emph{f}^{\;(l)}_{u,v}$, as shown in Eq.(4), is the diffractive connection between the $\emph{u}$-th metaunit locates at $(x_u,y_u)$ of the ($\emph{l}-1$)-th metaline and the $\emph{v}$-th metaunit locates at $(x_v,y_v)$ in the $\emph{l}$-th metaline, cos\,\emph{$\theta_{u,v}$}=$(x_u-x_v)/r_{u,v}$, $r_{u,v}=\sqrt{(x_u-x_v)^2+(y_u-y_v)^2}$ denotes the distance between the two metaunits, $\lambda$ is working wavelength, \emph{j} is imaginary unit, $\eta$ and $\Delta \psi$ are amplitude and phase coefficients, respectively. As for each metaline, the introduced phase modulations is modeled by a $\emph{V} \times \emph{V}$ diagonal matrix $\textbf{T}^{\,(l)}$, as expressed in Eq.(5):

\begin{equation}
\textbf{T}^{\,(l)}=
\left[ \begin{array}{ccccc}
{\rm exp}(j\Delta \phi^{(l)}_{1})  & 0 & \cdots & 0\\
0  & {\rm exp}(j\Delta \phi^{(l)}_{2}) & \cdots & 0\\
\vdots & \vdots &\ddots & \vdots\\
0  & 0 & \cdots & {\rm exp}(j\Delta \phi^{(l)}_{V})\\
\end{array} 
\right ]
\end{equation}

To proof the accuracy of the proposed model in Eq.(4) and Eq.(5), we evaluate the optical field of an OCU with finite-different time-domain (FDTD) method as shown in Fig.2(a). Three metalines with 10 metaunits for each are configured based on a standard SOI platform, the size of slab waveguide between metalines is $40 \rm um \times 15 \rm um$, the width and gap of slots are set to be 200nm and 500nm, the period of metaunit is 1.5um, and a laser source is split to 9 waveguides with working wavelength of 1550nm. We monitor the amplitude and phase response of the diffracted optical field in position A of Fig.2(a), who are well-agreed with the proposed analytical model in Eq.(4), as shown in Fig.2(b) and (c). Phase modulation of the metaline is also validated by monitoring the optical phase response at position B in Fig.2(d), with the incident light of a plane wave. Fig.2(e) shows a great match between the FDTD calculation and the analytical model in Eq.(5).

Consequently, we conclude the OCU model in Eq.(6), where \emph{M} is the layer number of OCU and $\textbf{R}_{OCU}$ is the response of the OCU when the input is a reshaped image patch matrix $\widehat{\emph{\textbf{I}}}$. Besides, the numbers of metaunits in \emph{M} metaline layers are all designed to be \emph{V}, which leads to $\textbf{F}^{(l+1)}$ and $\textbf{T}^{(l)}$ (\emph{l} = 1, 2, ..., \emph{M}-2) are matrices with size of $V \times V$. Specifically, $\textbf{F}^{(1)}$ is a $V \times H^2$ matrix since the $H^2$ waveguides are exploited and $\textbf{F}^{(M)}$ is a $2 \times V$ matrix since we only focus on the signals at two output ports.

\begin{equation}
\textbf{R}_{OCU}= \Big\{\textbf{F}^{(M)} \textbf{T}^{(M-1)}\Big[\prod_{l=1}^{M-2}(\textbf{F}^{\,(l+1)} \textbf{T}^{\,(l)})\Big]\textbf{F}^{(1)}\Big\} \cdot \widehat{\textbf{I}}
\end{equation}
Therefore, $\textbf{R}_{OCU}$ is a $2 \times G^2$ matrix with column of $\textbf{R}_1$ and $\textbf{R}_2$, which are $1 \times G^2$ vectors, and the corresponding response of balanced detection is described in Eq.(7) accordingly, where $\odot$ denotes a Hadamard product and $\kappa$ is a amplitude coefficient introduced by the photodetection:
\begin{equation}
\textbf{R}_{BPD}=\kappa\Big\{\Vert\textbf{R}_1\odot\textbf{R}_1^*\Vert-\Vert\textbf{R}_2\odot\textbf{R}_2^*\Vert\Big\}
\end{equation}

Furthermore, the OCU and balanced detection model in Eq.(6) and Eq.(7) can be abstracted as a feedforward neural network as illustrated in Fig.1(d), where the dense connections denotes diffractions and the single connections are phase modulations introduced by metalines. BPD's square-law detection performs as a nonlinear activation in the network since the phase-involved computing makes the network complex-valued \cite{hirose2003complex, ozdemir2011complex,scardapane2018complex}. 

Note that it is rarely possible to build a one-to-one mapping between the metaunit lengths and kernel value directly, because the phase modulation of metalines introduce complex-valued computations while the kernels are usually real-valued. However, the feedforward linear neural network nature of OCU model facilitates another approach to implement 2D convolution optically. \textbf{Structural re-parameterization} \cite{ding2021diverse,ding2021repvgg,ding2021resrep} (SRP) is a networking algorithm in deep learning, in which the original network structure can be substituted equivalently with another one to obtain same outputs, as illustrated in Fig.3. Here, we leverage this concept to make a regression between the diffractive feedforward neural network and 2D convolution. In another word, we train the network to learn how to perform 2D convolution instead of mapping the kernel value directly into metaunit lengths. More details are shown in following sections.

\section{Results}
In this section, we evaluate the performance of OCU in different aspects of deep learning. In subsection A, we present the basic idea of 2D optical convolution with the concept of SRP, and we demonstrate that the proposed OCU is capable of representing arbitrary real-valued $H \times H$ convolution kernel (in our following demos, we take $\emph{H}=3$) and therefore implementing basic image convolution optically. In subsection B, we take the OCU as a fundamental unit to build an oCNN, with which classification and regression applications of deep learning are carried out with remarkable performances.

\subsection{Optical convolution functioning}
As mentioned above, an OCU can not be mapped from a real-valued kernel directly since the phase modulation of metalines makes the OCU model a complex-valued feedforward neural network. Therefore, we need to train the OCU to "behave" as a real-valued convolution model with SRP method, which is referred as the training phase of OCU, and this idea is illustrated in Fig.4(a). We utilize a random pattern as the training set to make a convolution with a real-valued kernel, and the corresponding result is reshaped as a training label $\widehat{\textbf{R}}$. Then we apply the training set on the OCU model to get a feature vector $\textbf{R}_{BPD}$ and calculate a mean square error (MSE) loss $\mathbb{J}$ with the collected label. Through the iteration of backward propagation algorithm on our model, all the trainable parameters are updated to minimize loss and the OCU is evolved to the targeting real-valued kernel, as shown in Eq.(8) and Eq.(9), where $\Delta \textbf{$\Phi$}$ is metaline-introduced phase and it is also the trainable parameter of OCU. Accordingly, images can be convolved with the well-trained OCU, and we term this process as an inference phase, as presented in Fig.4(b).

\begin{equation}
\mathbb{J} = \frac{1}{2} \cdot \sum_{i=1}^{G^2}\Vert \ \textbf{R}_{BPD}(\Delta \textbf{$\Phi$})[i] - \widehat{\textbf{R}}[i] \ \Vert ^2\\
\end{equation}

\begin{equation}
\Delta \textbf{$\Phi$}^* = \mathop{\arg\min}_{\Delta \textbf{$\Phi$}} \ \ \ \mathbb{J}(\Delta \textbf{$\Phi$})
\end{equation}

For proof-of-concept, a $128 \times 128$ random pattern (the OCU's performance gets almost no improvement with a random pattern that is larger than $128 \times 128$) and 8 unique real-valued $3 \times 3$ convolution kernels are exploited to generate training labels and a $256 \times 256$ gray scale image is utilized to test the OCUs' performance, as shown in Fig.5. In this case, we use 3 layers of metalines in OCU with $L_1$ = 75um and $L_2$ = 300um, each metaline consists of 50 metaunits with $w_1$ = 200nm, $g$ = 500nm and $p$ = 1.5um, the number of input waveguides are set to be 9 according to the size of the utilized real-valued kernel. The training and inference process of OCU are conducted with Tensorflow2.4.1 framework. From Fig.5 we can see great matches between the ground truths generated by real-valued kernels and the outputs generated by OCUs, and the average MSE between the two can be calculated as 0.0405, indicating that the OCU can response as a real-valued convolution kernel with remarkable performance.

\subsection{Optical convolutional neural network (oCNN)}
With OCU as the basic unit for feature extraction, more sophisticated architectures can be carried out efficiently to interpret the hidden mysteries in higher dimensional tensors. In this section, we build an optical convolutional neural network (oCNN) to implement tasks in two important topics of deep learning: classification and regression. 

\subsubsection{Image classification}
Fig.6 shows the basic architecture of oCNN for image classifications. Images with size of $N \times N \times C$ are firstly flatten into $C$ groups of patches and concatenated as a data batch with size of $G^2 \times C \cdot H^2$ according to the kernel size $H$, and then loaded to a modulator array with totally $C \cdot H^2$ modulators in parallel. Here, $C$ denotes the image channel number, and $N$, $G$, $H$ are already defined in the principle section. The modulated data batch is copied $q$ times and splitted to $q$ optical convolution kernels (OCKs) by means of optical routing. Each OCK consists of $C$ OCUs corresponding to $C$ data batch channels, and the $n$-th channel of the data batch is convovled by the $n$-th OCU in each OCK, where n = 1,2,...$C$. Balanced photodetection is utilized after each OCU to give a sub-feature map FM$_{mn}$ with size of $G \times G$, where m = 1,2,...$q$, and all $C$ sub-feature maps in a OCK are summed up to generate a final feature map FM$_{m}$. For convenience, we term this process as optical convolution layer (OCL) as denoted inside the blue dashed box of Fig.5. After OCL, the feature maps are further downsampled by the pooling layer to form more abstracted information. Multiple OCLs and pooling layers can be exploited to establish deeper networks when the distribution of tensors (herein this case, images) is more complicated. At last, the highly extracted output tensors are flatten and sent to a small but fully connected (FC) neural network to play the final classifications.

We demonstrate the oCNN classification architecture on gray scale image dataset Fashion-MNIST and colored image dataset CIFAR-4 that selected from the widely used CIFAR-10 with much more complex data distribution. We visualize the two datasets with t-distributed stochastic neighbor embedding (t-SNE) method in a 2D plane as shown in Fig.7(d). For Fashion-MNIST, we use 4 OCKs to compose an OCL for feature extraction and 3 cascaded FC layers to give the final classification, assisted with the loss of cross entropy. Here in this case, each OCK only has one OCU since gray-scale images have only one channel, and each OCU performs as a $3 \times 3$ convolution. We use 60000 samples as the training set and 10000 samples as the test set, after 500 epochs of iterations, the loss of both training set and test set are converged and the accuracy of test set is stable at 91.63\%, as given in Fig.7(a) attached with a confusion matrix. For CIFAR-4 dataset, similar method is leveraged: an OCL with 16 OCKs are carried out with each OCK consists of 3 OCUs according to R, G, B channels of the image, and then 3 FC layers are applied after the OCL. And each OCU also performs as a $3 \times 3$ convolution. Here, 20000 samples are used as training set and another 4000 samples as test set, the iteration epoch is set as 500, we also use cross entropy as the loss function. After iterations of training, the classification accuracy is stable at 86.25\% as shown in Fig.7(b), and the corresponding confusion matrix is also presented. The OCU's parameter we use here is as same as the settings in sub-section A. Furthermore, we also evaluate the performances of electrical neural networks (denoted as E-net) with the same architecture as optical ones in both two datasets, and the results show that the proposed oCNN outperforms E-net with accuracy boosts of 1.14\% for Fashion-MNIST and 1.75\% for CIFAR-4. 

We also evaluate the classification performance of the oCNN respect to two main phyiscal paramaters of the OCU: the number of metaunit per layer and the number of the exploited metaline layer, as shown in Fig.7(c). In the left of Fig.7(c), 3 metaline layers are used with the number of metaunit per layer varied from 10 to 70, and the result shows that increasing the metaunit numbers gives accuracy improvements for both datasets, but the task for CIFAR-4 has a more significant boost of 6.73\% than Fashion-MNIST of 2.92\% since the former one has a more complex data structure than the latter, therefore it is more sensitive to model complexity. In the right of Fig.7(c), 50 metaunits are used for each metaline layer, and the result indicates that increasing the layer of metaline also gives a positive response on test accuracy for both datasets, with accuracy improvments of 1.45\% and 1.05\%, respectively. To conclude, the oCNN can further improve its performance by increasing the metaunit density of the OCU, and adding more metaunits per layer is a more efficient way than adding more layers of metaline to achieve this goal.

\subsubsection{Image denoising}
Image denoising is a classical and crucial technology that has been widely applied for high performance machine vision \cite{goyal2020image, tian2020deep}. The goal is to recover a clean image $\textbf{X}$ from a noisy one $\textbf{Y}$, and the model can be written as $\textbf{Y} = \textbf{X} + \textbf{N}$, where in general $\textbf{N}$ is assumed to be a additive Gaussian noise (AWGN). Here, we refer the famous feed-forward denoising convolutional neural network (DnCNN) \cite{zhang2017beyond} to build its optical fashion, termed as optical denoising convolutional neural network (oDnCNN), to demonstrate the feasibility of the proposed OCU in deep learning regression.

Fig.8(a) shows the basic architecture of oDnCNN. The oDnCNN includes three different parts: (i) Input layer: OCL with $q_1$ OCKs is utilized, whose details are presented in Fig.6. Each OCK consists of $C_{in}$ OCUs which perform  $3 \times 3 \times C_{in}$ 2D convolutions, where $C_{in} = 1$ for gray scale images and $C_{in} = 3$ for colored images. Then ReLUs are utilized for nonlinear activation. (ii) Middle layer: OCL with $q_2$ OCKs is exploited, for the first middle layer $q_1$ OCUs are used in each OCK and for the rest of the middle layers the number is $q_2$. ReLUs are also used as nonlinearity and batch normalization is added between OCL and ReLU. (iii) Output layer: only one OCL with one OCK is leveraged which has $q_2$ OCUs.

With this architecture, basic Gaussian denoising with known noise level $\sigma$ is performed. We follow \cite{chen2016trainable} to use 400 gray scale images with size of $180 \times 180$ to train the oDnCNN and cropped them into $128 \times 1600$ patches with patch size of $40 \times 40$. For test images, we use the classic Set12 dataset that contains 12 gray images with size of $256 \times 256$. Three noise levels, i.e. $\sigma$ = 10, 15, 20, are considerd to train the oDnCNN, and also applied to the test images. The oDnCNN we apply for this demonstration includes one input layer, one middle layer and one output layer, among which 8 OCKs are exploited for both input and middle layer respectively, and 1 OCK for output layer. Similar physical parameters of OCUs in the OCKs are set as the ones in sub-section A, the only difference is we use only 2 layers of metaline in this denoising demo. Fig.8(b) shows the denoising results under test images with noise level $\sigma = 20$. We evaluate the average peak signal-to-noise ratio (PSNR) for each image before and after the oDnCNN's denoising as 22.10dB and 27.72dB, posing a 5.62dB improvement of image quality, and details in red boxes also show that clearer textures and edges are obtained at the oDnCNN's output. More demonstrations are carried out for noise level $\sigma = 10, 15$ and the performances of E-net are also evaluated, as presented in Table.1. The results reveal that the oDnCNN provides 3.57dB and 4.78dB improvements of image quality for $\sigma = 10$ and $\sigma = 15$, which is comparable with the E-net's performance. Our demonstrations are limited by the computation power of the utilized server, and the overall performance can be further improved by increasing metaunit density of the OCUs. 

\begin{table}[h]
    \centering
    \begin{tabular}{cccc}
    \toprule 
    \textbf{Noise level} & \textbf{Noisy (dB)} & \textbf{oDnCNN (dB)} & \textbf{E-net (dB)}\\
    \midrule  
    $\sigma = 10$ & 28.13 & 31.70 & 30.90\\
    $\sigma = 15$ & 24.61 & 29.39 & 29.53\\
    $\sigma = 20$ & 22.10 & 27.72 & 27.74\\
    \bottomrule 
    \end{tabular}
    \caption{Performance comparisons of the proposed oDnCNN and E-net in average PSNR, with noise level $\sigma $ = 10, 15, and 20.}
\end{table}

\section{Discussion}
\subsection{Computation speed and power consumption}
The operation number of a 2D convolution composes the production part and accumulation part, which can be addressed by the kernel size $H$ as the first equation in Eq.(10) shows. Consequently, for a convolution kernel in CNN, the operation number can be further scaled by input channel $C$, shown in the second equation in Eq.(10). Here, $O_{conv}$ and $O_{kernel}$ denote operation numbers of a 2D convolution and a convolution kernel in a CNN.

\begin{equation}
\begin{split}
    O_{conv} &= 2 \cdot H^2 -1\\
    O_{kernel} &= C \cdot O_{conv}\\
\end{split}
\end{equation}

Consequently, the computing speed of an OCU can be calculated by the operation number $O_{conv}$ and modulation speed $r$ of OMA, and the speed of an OCK with $C$ input channels can be also acquired, by evaluating the number of operations per second (OPS). The calculations are presented in Eq.(11), where $S_{ocu}$ and $S_{ock}$ represents the computation speed of OCU and OCK, respectively. 

\begin{equation}
\begin{split}
    S_{ocu} &= O_{conv} \cdot r \quad OPS\\
    S_{ock} &= C \cdot S_{ocu} \quad OPS\\
\end{split}
\end{equation}

From Eq.(11), we can see that the computation speed of the OCU or OCK is largely dependent with modulation speed of OMA. Meanwhile, high speed integrated modulator has been received considerable interests at present, both in terms of new device structure or new materials, and the relative industries are also going to be mature \cite{rahim2021taking}. Assume that the modulation speed is 100 GBaud per modulator, for an OCU performing $3 \times 3$ optical convolutions, the computation speed can be calculated as $(2 \times 3 \times 3 - 1) \times 100 $ = 1.7 TOPS. For instance, in the demonstration in the last section, 16 OCKs are utilized to classify the CIFAR-4 dataset who contains 3 channels for each image, therefore the total computation speed of the OCL can be addressed as $3 \times 1.7 \times 16$ = 81.6 TOPS.

Because the calculations of OCU are all passive, its power consumption mainly comes from the data loading and photodetection process. Schemes of photonics modulator with small driving voltage \cite{samani2015low,baehr2012ultralow,sakib2021high} have been proposed recently to provide low power consumption, and integrated photodetectors \cite{liu2021silicon, bie2017mote2} are also investigated with negligible energy consumed. Therefore, the total power of an OCU with equivalent kernel size of $H$ can be calculated as Eq.(12), where $E_{ocu}$, $E_{dri}$ and $E_{dec}$ are the energy consumptions of OCU, data driving and detection respectively, $E_{b}$ is the utilized modulator's energy consumption, $P_{d}$ is the power of photodetector, $B$ denotes symbol or pixel number and $D$ is the symbol precision. Assume that a 100 Gbaud modulator and a balanced photodetecor with energy and power consumption of 100fj/bit and 100mw are used, for a 4K image with more than 8 million pixels and 8-bit depth for each, the total energy consumed by a $3 \times 3$ optical convolution can be calculated as $(3 \times 3) \times (8 \times 10^6 \times 8 \times 3 \times 100 \times 10^{-15}) + 0.1 \times (8 \times 10^{-6}/(100 \times 10^9))  = 1.808 \times 10^{-4}$ J.

\begin{equation}
\begin{split}
    E_{mod} & = H^2 \cdot (B \cdot D \cdot C \cdot E_b)\\
    E_{det} & = P_d \cdot (B/r)\\
    E_{ocu} & = E_{mod} + E_{det}\\
\end{split}
\end{equation}

\subsection{Networking with optical tensor core (OTC)}
Today's cutting-edge AI systems are facing double test of computation forces and energy cost in performing data-intensive applications\cite{lecun20191}, models like ResNet50\cite{he2016deep} and VGG16\cite{simonyan2014very} are power-hungry in processing high dimensional tensors such as images and videos, molecular structures, time-serial signals, languages, etc., especially when the semiconductor fabrication process approaches its limitation\cite{schaller1997moore}. Edge computing\cite{shi2016edge, mao2017survey, satyanarayanan2017emergence}, a distributive computing paradigm, is proposed to process data near its source to mitigate bandwidth wall and further improve computation efficiency, which requires computing hardware has low run-time energy consumption and short computing latency. Compute-in-memory (CIM) \cite{ielmini2018memory, sebastian2020memory, verma2019memory} receives considerable attentions in recent years since it avoids long time latency in data movement and reduces intermediate computations, showing a great potential as AI edge processor. However, reloading large-scale weights repeatedly from DRAM to local memories also weakens energy efficiency significantly. Notably, the proposed OCU can be regarded as a natural CIM architecture because the computations are performed with the optical flow connecting the inputs and the outputs with the speed of the light, and more importantly, its weights are fixed at the metaunits and therefore the data loading process is eliminated.

Consequently, from a higher perspective, we consider a general networking method with multiple OCUs and optoelectrical interfaces, by leveraging the idea of network rebranching \cite{chen2022yoloc}, to build an optoelectrical CIM architecture, as shown in Fig.9. The idea of rebranching is to decompose the model mathematically into two parts: trunk and branch, and by fixing the major parameters in trunk and altering the minor ones in branch, the network can be programmed with very low energy consumption. For the trunk part, who is responsible for major computations of the model, has fixed weights provided optically, referred as optical tensor core (OTC): Laser bank is exploited as the information carrier and routed by optical I/O to multiple optical tensor units (OTUs), and tensor data in the DRAM is loaded into OTUs by high speed drivers. The OTUs which contains modulator array, OCUs and balanced PD array manipulate tensor convolutions passively, and the calculation results are read out by DSPs. As for branch part, who is a programmable lightweight electrical network, is responsible for reconfiguring the whole model with negligible computations. With this structure, big models can be performed with speed of TOPS level but almost no power is ever consumed, and time latency is also shorten since much fewer weights are reloaded from the DRAM. This scheme is promising for future photonics AI edge computing.

Technologies for the implementation of OTC are quite mature in these days. DFB laser array \cite{mukherjee2000wdm, buckley2018wdm} can be applied as the laser bank which has been widely used in commercial optical communication systems, and on-chip optical frequency comb \cite{chang2022integrated,kippenberg2011microresonator,chembo2016kerr} can provide even more compact and efficient source supply with Kerr effect in silicon nitride waveguide. Integrated optical routing schemes are proposed recently based on MZI network\cite{shoji2010low,lu201616,qiao201732}, ring modulators \cite{dong2007all,wen2011all,lee2008all, sherwood2008optical} and MEMS \cite{han2018large, kwon2018128,hwang2017flip}, with very low insertion loss and flexible topology. Integrated modulators with ultrahigh bandwidth and low power consumption are also investigated intensively based on MZ \cite{liao2005high} and ring \cite{sakib2021high} structure, with diverse material platforms including SOI \cite{rahim2021taking}, lithium niobate \cite{wang2018integrated} and indium phosphide\cite{ogiso201980}. High speed photodetectors with high sensitivity, low noise and low dark current based on either silicon or III-V materials have also been studied and massively produced for optical communications \cite{zhao2017high} and microwave photonics \cite{malyshev2004state} industries. The existing commercial silicon photonics foundries \cite{lim2013review, siew2021review} are capable of fabricating metasurfaces with minimum linewidth smaller than 180nm via universal semiconductor techniques, showing a potential for future pipeline-based production of the proposed OCU.

\section{Conclusion}
In this work, we proposed an optical convolution architecture, OCU, with light diffraction on 1D metasurface to process large scale tensor information. We demonstrate that our scheme is capable of performing any real-valued 2D convolution by using the concept of structural re-parameterization. We then apply the OCU as a computation unit to build a convolutional neural network optically, implementing classification and regression tasks with extraordinary performances. The proposed scheme shows advantages in either computation speed or power consumption, posing a novel networking methodology of large-scaled while lightweight deep learning hardware frameworks.

\section{Funding}
This work was supported by the National Natural Science Foundation of China (NSFC) (62135009).

\bibliographystyle{unsrt}

\bibliography{sample}

\newpage

\begin{figure*}[htb]
\centering\includegraphics[width=15.5cm]{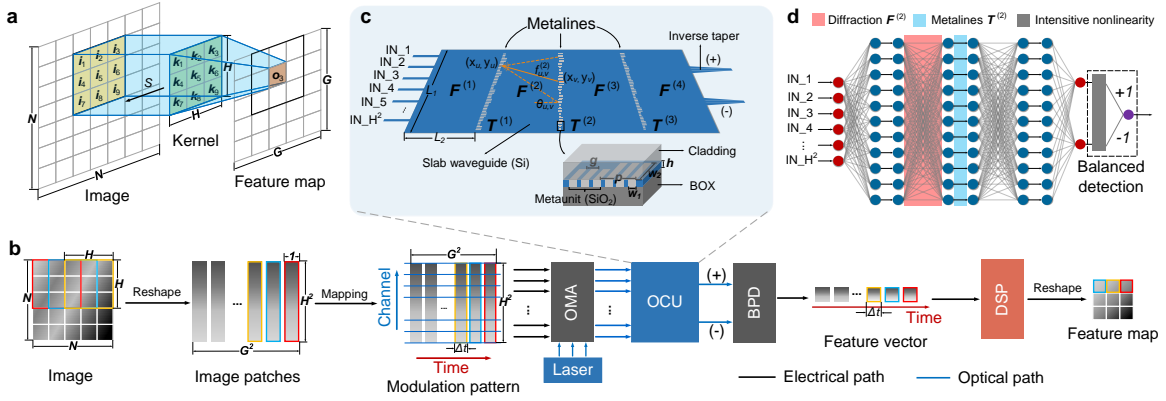}
\caption{\footnotesize Principle of optical image convolution based on OCU. (\textbf{a}) Operation principle of 2D convolution. A fixed kernel with size of $H\times H$ slides over the image with size of $N\times N$ by stride of $S$ and does weighted addition with the image patches that cover by the kernel, resulting an extracted feature map with size of $G\times G$, where $G = \lfloor (N - H)/S + 1 \rfloor$. (\textbf{b}) Optical image convolution architecture with OCU. An image is firstly flatten into patches according to the kernel size and sliding stride, and then mapped into a modulation pattern confined with time and channel number, which modulates a coherent laser via a modulation array. The modulated light is sent to OCU to perform optical convolution, whose positive and negative results are subtracted by a balanced photodetector and reshaped by a DSP to form a new feature map. OMA, optical modulator array; BPD, balanced photodetector; DSP, digital signal processor. (\textbf{c}) The details of OCU. $H^2$ waveguides are used to sent laser signal into a silicon slab waveguide with size of $L_1 \times L_2$, and layers of metaline are exploited successively with layer of gap of $L_2$ which are composed by well-arranged metaunits. Three identical silica slots with sizes of $w_1 \times w_2 \times h$ are used to compose one metaunit with gap of $g$, and the period of metaunits is $p$ The phase modulation is implemented by varying $w_2$. The transfer function of the diffraction in slab waveguide and phase modulation of metalines are denoted as $\textbf{F}$ and $\textbf{T}$. (\textbf{d}) The feedforward neural network abstracted from the OCU model. Red and blue boxes denote dffractions and phase modulations of metalines, gray box represents a intensitive nonlinear activation of complex-valued neural networks introduced by photodetection.}
\label{fig:1}
\end{figure*}

\begin{figure}[!ht]
\centering\includegraphics[width=8.8cm]{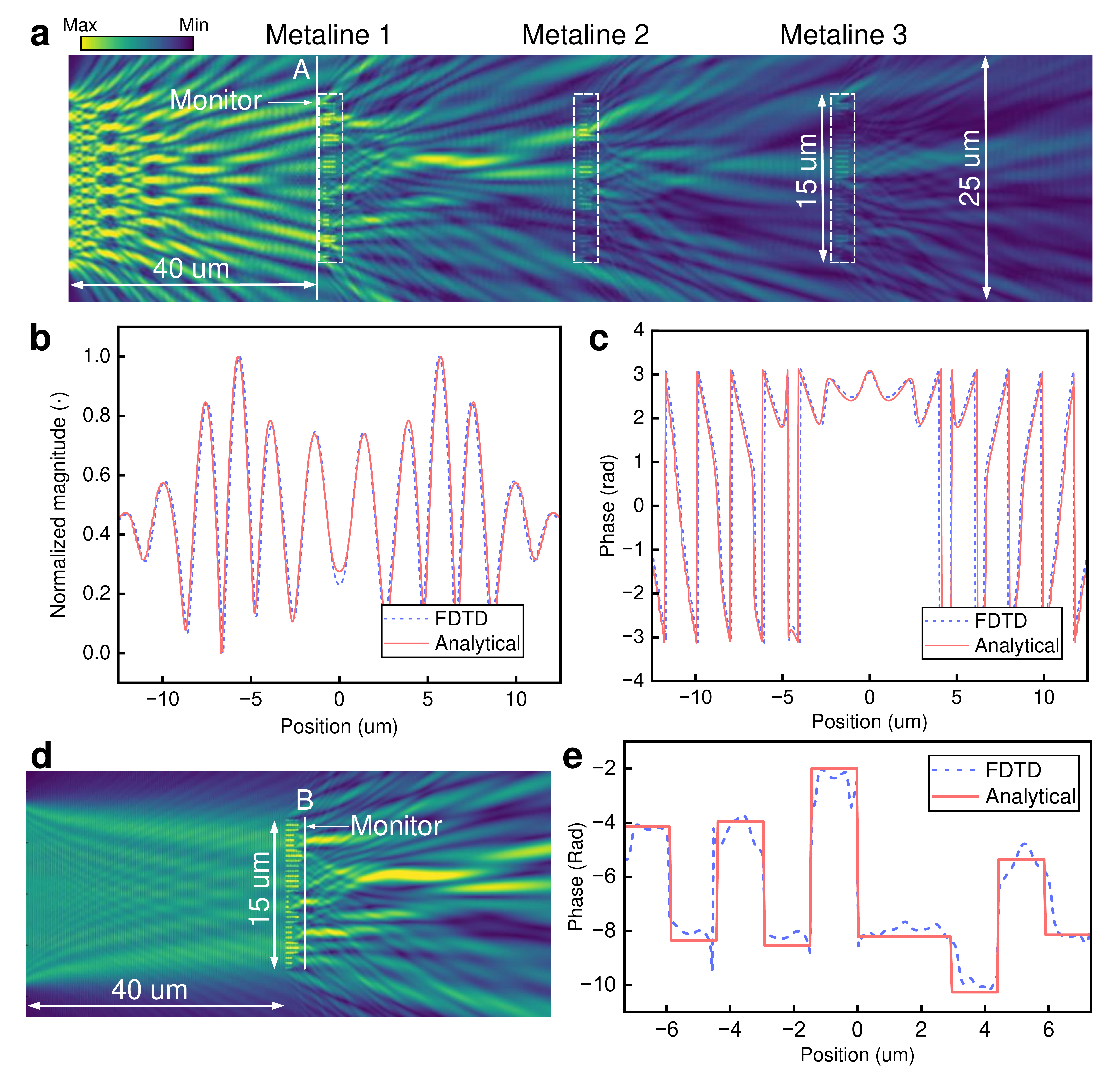}
\caption{\footnotesize (\textbf{a}) The Optical field of the OCU evaluated by FDTD method. A monitor is set at position A to receive the optical field of the incident light. (\textbf{b}) Magnitude and (\textbf{c}) phase response of the optical field at position A (red solid curve) match well with the analytical model (purple dash curve) in Eq.(5). (\textbf{d}) Optical field of the the metaline with incident light of a plane wave. A monitor is set behind the metaline at position B to obtain its phase response. (\textbf{e}) The analytical model (perple dash curve) of Eq.(6) fits well with the FDTD calculation (red solid curve).}
\label{fig:2}
\end{figure}

\begin{figure}[!ht]
\centering\includegraphics[width=8.6cm]{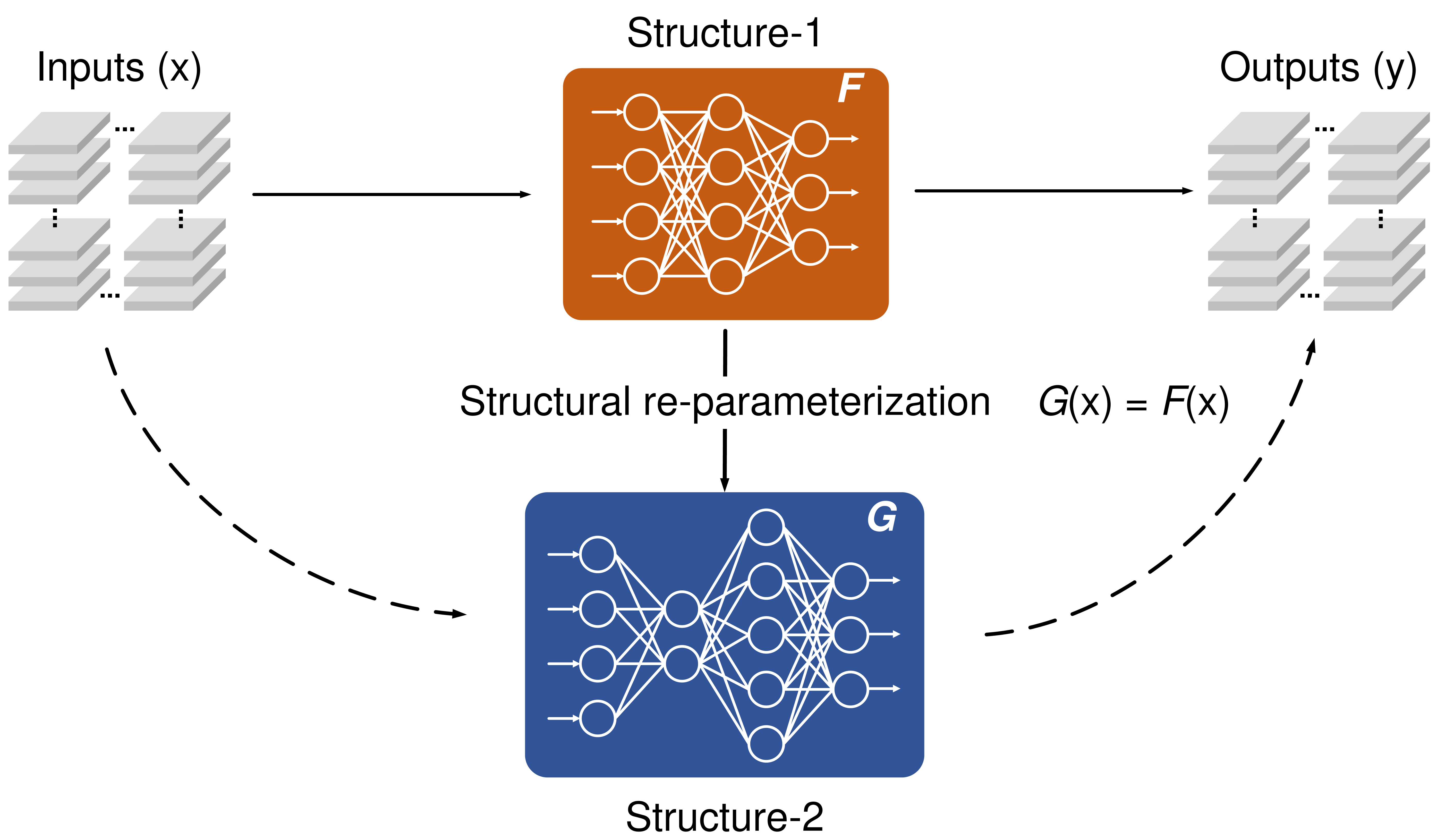}
\caption{\footnotesize Concept of structural re-parameterization in deep learning. Network structure-1 has a transfer function of $F$, which can be substituted equivalently by a different network structure-2 whose transfer function is $G$. Accordingly, both structures have the same outputs $y$ under the same inputs of $x$. }
\label{fig:3}
\end{figure}

\begin{figure*}[!ht]
\centering\includegraphics[width=15.5cm]{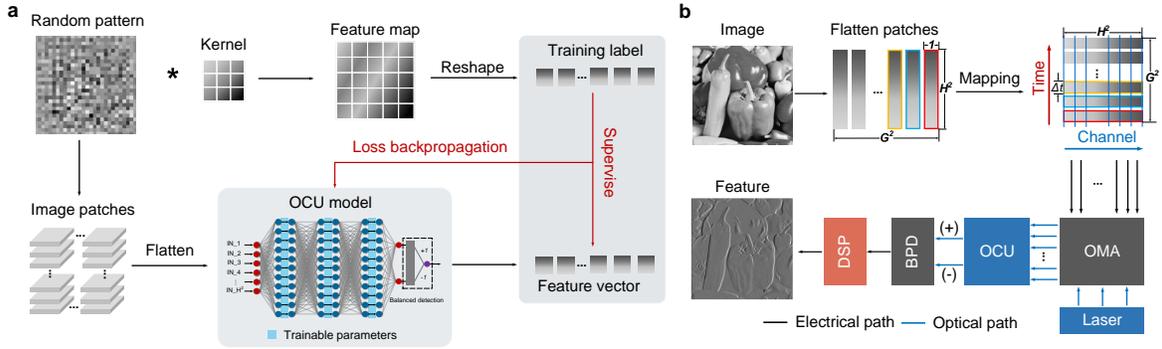}
\caption{\footnotesize The training and the inference phase of an OCU to perform real-valued optical convolution with the idea of SRP in deep learning. (\textbf{a}) A $128 \times 128$ random pattern is utilized to generate a training pair for OCU model. The output feature vector of OCU model is supervised by the training label with the input of flatten image patches that decomposed from the random pattern. (\textbf{b}) A $256 \times 256$ gray scale image is reshaped to flatten patches and sent to the well-trained OCU to perform a real-valued convolution. OMA, modulatior array; BPD, balanced photodetector; DSP, digital signal processor.}
\label{fig:4}
\end{figure*}

\begin{figure}[!ht]
\centering\includegraphics[width=8.3cm]{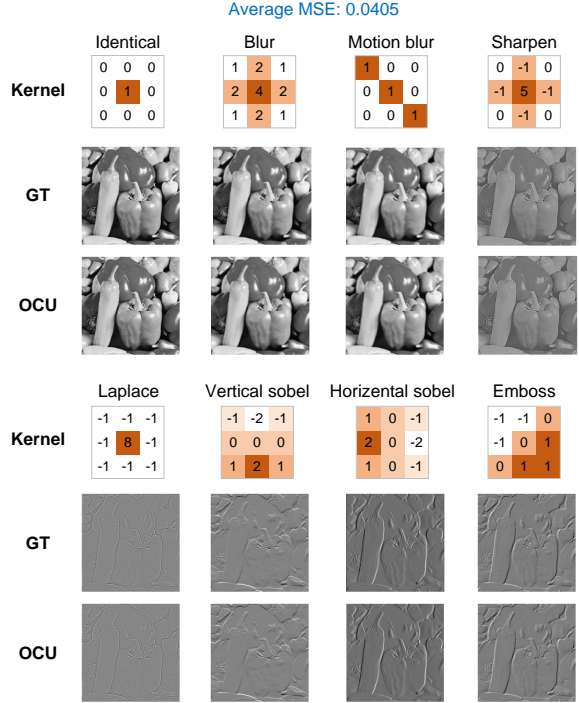}
\caption{\footnotesize Well-agreed convolution results between the ground truths and the outputs of OCUs with 8 unique real-valued convolution kernels, with average MSE of 0.0405. GT, ground truth.}
\label{fig:5}
\end{figure}

\begin{figure*}[!ht]
\centering\includegraphics[width=15.5cm]{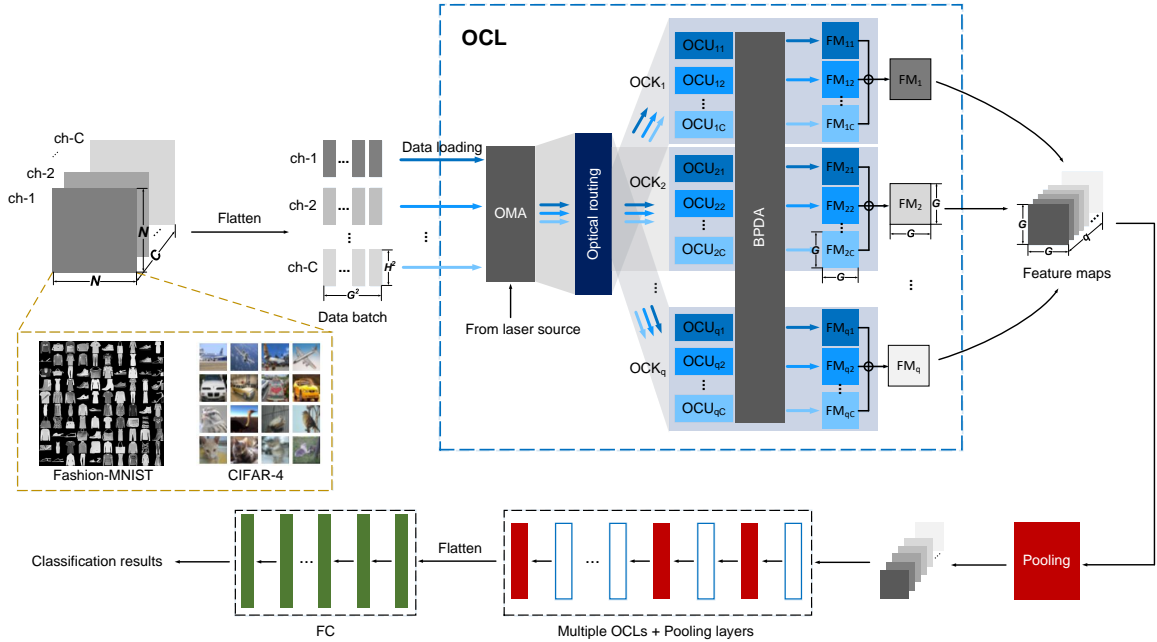}
\caption{\footnotesize Architecture of oCNN for image classification. Images with size of $N \times N \times C$ are firstly flatten into $C$ groups of patches and concatenated as a data batch with size of $G^2 \times C \cdot H^2$ according to the kernel size $H$, and then loaded to a modulator array with totally $C \cdot H^2$ modulators in parallel. The modulated signal is splitted to $q$ OCKs by optical router, each of which contains $C$ OCUs to generate $C$ sub-feature maps, and then all the sub-feature maps of each OCK are summed up to form a final feature map. We refer this process as an OCL denoted by the blue dash box. After OCL, the feature maps are further downsampled by pooling layer, and multiple OCLs and pooling layers can be utilized to build deeper networks to manipulate more complicated tasks. A small scale fully connected layer is used to give the final classification results. OMA, optical modulator array; OCK, optical convolution kernel; BPDA, balanced photodetector array; FM, feature map; FC, fully connected layer.}
\label{fig:6}
\end{figure*}

\begin{figure*}[!ht]
\centering\includegraphics[width=15.5cm]{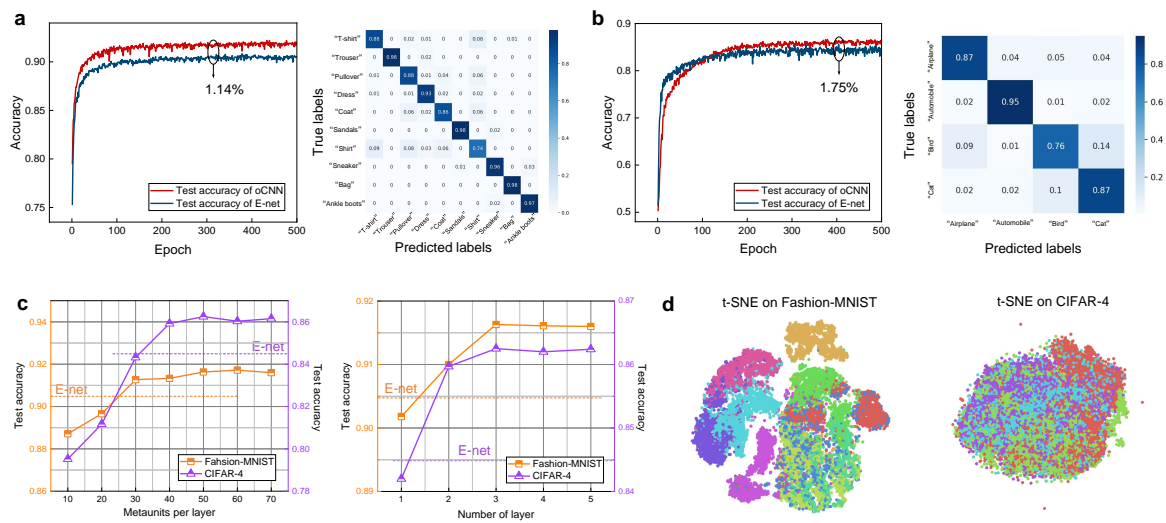}
\caption{\footnotesize Classification results of oCNNs for (\textbf{a}) Fashion-MNIST and (\textbf{b}) CIFAR-4 datasets. Accuracy of 91.63\% and 86.25\% are obtained with oCNNs for the corresponding two datasets, which outperform their electrial counterparts with 1.14\% and 1.75\% respectively. (\textbf{c}) Classification performance evaluations on both datasets respect to two main physical parameters of OCU: the number of metaunit per layer and the number of the exploited metaline layer. (\textbf{d}) 2D visualizations of the two applied datasets with t-distributed stochastic neighbor embedding (t-SNE) method.}
\label{fig:7}
\end{figure*}

\begin{figure*}[!ht]
\centering\includegraphics[width=15.5cm]{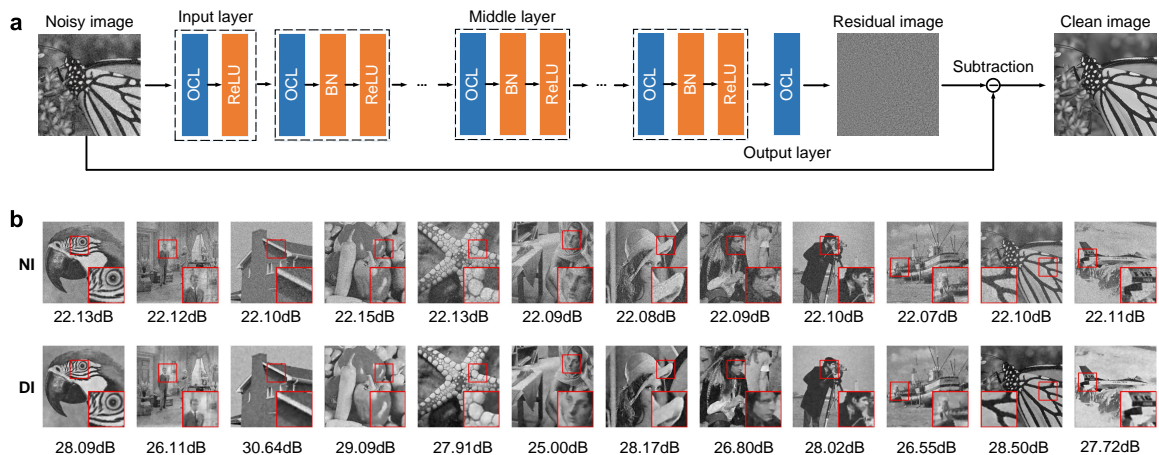}
\caption{\footnotesize (\textbf{a}) Architecture of the proposed oDnCNN. A Gaussian noisy image with a known noise level is firstly flatten and modulated into lightwave and then sent to oDnCNN, which composes three parts: input layer with OCL and ReLU, middle layer with an extra batch normalization between the two, and output layer with only an OCL. After the oDnCNN, a residual image is obtained which is the extracted noise. By subtracting the noisy image with the extracted residual one, the clean image can be aquired. OCL, optical convolution layer; ReLU, rectified linear units; BN, batch normalization. (\textbf{b}) The denoised result of Set12 dataset leveraged by the proposed oDnCNN with noise level $\sigma = 20$, giving much clearer textures and edges as the details shows in red boxes. In this case, the average PSNR of the denoised images is 27.02dB, compared with 22.10dB of the noisy ones. NI, noisy images; DI, denoised images.}
\label{fig:8}
\end{figure*}

\begin{figure*}[!ht]
\centering\includegraphics[width=15.5cm]{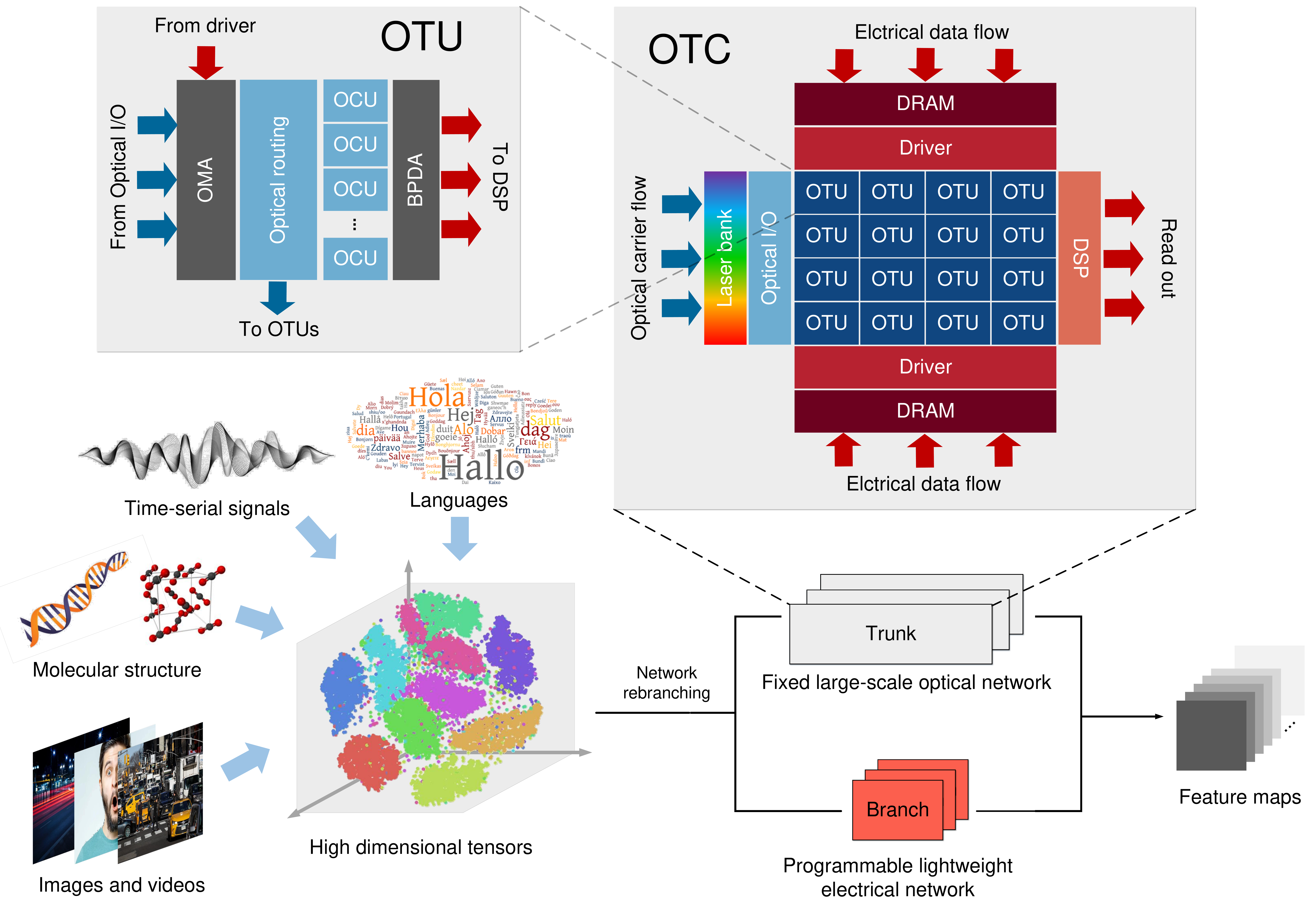}
\caption{\footnotesize Highly efficient optical deep learning framework with network rebranching and optical tensor core. Deep learning models are decomposed mathematically into two parts: trunk and branch, who carry the major and minor calculations of the model respectively. The trunk part is computed by optical tensor core with fixed weights and branch part is performed by a lightweight electrical network to reconfigure the model. OTC, optical tensor core; OTU, optical tensor unit.}
\label{fig:9}
\end{figure*}

\end{document}